%% file: main_paper.tex
\documentclass[sigconf]{acmart}
\usepackage[capitalize,noabbrev]{cleveref}

\AtBeginDocument{%
  }

\copyrightyear{2024}
\acmYear{2024}
\setcopyright{rightsretained}
\acmConference[WSDM '24]{Proceedings of the 17th ACM International Conference on Web Search and Data Mining}{March 4--8, 2024}{Merida, Mexico}
\acmBooktitle{Proceedings of the 17th ACM International Conference on Web Search and Data Mining (WSDM '24), March 4--8, 2024, Merida, Mexico}\acmDOI{10.1145/3616855.3635819}
\acmISBN{979-8-4007-0371-3/24/03}

\makeatletter
\gdef\@copyrightpermission{
  \begin{minipage}{0.3\columnwidth}
   \href{https://creativecommons.org/licenses/by/4.0/}{\includegraphics[width=0.90\textwidth]{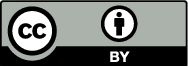}}
  \end{minipage}\hfill
  \begin{minipage}{0.7\columnwidth}
   \href{https://creativecommons.org/licenses/by/4.0/}{This work is licensed under a Creative Commons Attribution International 4.0 License.}
  \end{minipage}
  \vspace{5pt}
}
\makeatother

\input{macros.tex}

\begin{document}

\settopmatter{authorsperrow=4}

\title{Ranking with Long-Term Constraints}

\author{Kiant\'e Brantley}
\affiliation{%
  \institution{Cornell University}
  \city{Ithaca}
  \state{NY}
  \country{USA}
}
\email{kdb82@cornell.edu}

\author{Zhichong Fang}
\affiliation{%
  \institution{Cornell University}
  \city{Ithaca}
  \state{NY}
  \country{USA}
}
\email{zf94@cornell.edu }

\author{Sarah Dean}
\affiliation{%
  \institution{Cornell University}
  \city{Ithaca}
  \state{NY}
  \country{USA}
}
\email{sdean@cornell.edu}

\author{Thorsten Joachims}
\affiliation{%
  \institution{Cornell University}
  \city{Ithaca}
  \state{NY}
  \country{USA}
}
\email{tj@cs.cornell.edu}

\begin{abstract}
The feedback that users provide through their choices (e.g., clicks, purchases) is one of the most common types of data readily available for training search and recommendation algorithms. However, myopically training systems based on choice data may only improve short-term engagement, but not the long-term sustainability of the platform and the long-term benefits to its users, content providers, and other stakeholders. In this paper, we thus develop a new framework in which decision makers (e.g., platform operators, regulators, users) can express long-term goals for the behavior of the platform  (e.g., fairness, revenue distribution, legal requirements). These goals take the form of exposure or impact targets that go well beyond individual sessions, and we provide new control-based algorithms to achieve these goals. In particular, the controllers are designed to achieve the stated long-term goals with minimum impact on short-term engagement. Beyond the principled theoretical derivation of the controllers, we evaluate the algorithms on both synthetic and real-world data. While all controllers perform well, we find that they provide interesting trade-offs in efficiency, robustness, and the ability to plan ahead. 
\end{abstract}



\maketitle

\section{Introduction} \label{sec:intro}

Optimizing search and recommendation platforms based on feedback that users provide through their choices (e.g., clicks, purchases) has led to great improvements in ranking quality. However, myopically training systems based on choice data may only improve short-term engagement, but not the long-term sustainability of the platform and the long-term benefits to its users, content providers, and other stakeholders \cite{Hohnhold/etal/15}. In particular, platforms operate as part of a complex socio-technical system, and many have argued how such AI systems can amplify misinformation \cite{FernandezB20}, harm supply through rich-get-richer dynamics \cite{Mehrotra/etal/18}, incentivize spam \cite{Nitin/Bing/07}, or perpetuate human biases \cite{Mansoury/etal/20}.

In this complex space of problems and competing interests, we argue that improved tools for explicitly steering the long-term dynamics of the platform are needed. 
These tools should enable decision-makers to specify long-term goals for the search and recommendation algorithms beyond short-term engagement maximization. 
While approaches using reinforcement learning have the potential to directly optimize long-term goals, 
it is challenging to apply them to complex and large-scale information retrieval settings~\citep{afsar2022reinforcement,lin2023survey}.
End-to-end frameworks obscure important decision points~\cite{gilbert2022reward}, leading to problems like a lack of reproducibility~\citep{engstrom2020implementation, henderson2018deep, andrychowicz2020matters}, reward hacking~\citep{ng1999policy, ng2000algorithms,everitt2017reinforcement, skalse2022defining, pan2022effects}, and user tampering and manipulation~\citep{carroll2021estimating,krueger2020hidden,evans2021user,everitt2021reward}.
Instead, we argue that providing designers with a novel {\em macroscopic} view will enable strategic reasoning about long-term platform dynamics and that it will enable new tools for steering the platform.
The key algorithmic challenge lies in bridging the gap between long-term goals at a macro-level that span many requests, and the micro-level goal of maximizing engagement for each individual request.

In this paper, we develop a new class of macro-level interventions for steering the long-term dynamics of AI platforms, as well as the mechanisms for optimally executing these macro-level interventions. In our framework, further described in Sections~
\ref{sec:framework}  and~\ref{sec:macro_micro_problem}, macro-level interventions take the form of exposure or impact targets over substantial periods of time (e.g., days, weeks, months). 
The macro-level interventions can come from various decision-makers, including the users themselves (e.g., "I want to buy at least 30\% local products next month" on an e-commerce platform), the platform operator (e.g., "promote local music communities by serving at least 50\% local artists on average" on the Localify! music platform \cite{Localify}), or regulators (e.g., the recent settlement between Meta and the Department of Housing and Urban Development (HUD) \cite{facebooksettlement/22} that requires Meta to ensure that each housing-related ad is shown with demographic parity to all protected groups over the ad's lifetime). 
All such macro-level goals steer the behavior of the system over the course of many requests. This creates complex interactions between individual requests, their short-term metrics, and the long-term goals \cite{Morgenstern/21}.

We address 
the key technical problem of designing algorithms which 
break down macro-level goals
into a sequence of individual rankings that least hurt the micro-level metric (e.g., engagement). 
We view these algorithms as controllers which drive the value of macro-level metrics towards specified targets while responding to incoming requests in real time.
In Section~\ref{sec:controllers}, we rigorously derive three controllers.
The first
is a baseline approach that satisfies the macro-level goals at a high cost to the micro-level utility.
The second
enables a finer trade-off between macro- and micro-level objectives. 
The final controller
incorporates planning to handle requests coming from non-stationary distributions with temporal patterns. 
To clarify the design of these controllers and their affordances and limitations, we make interesting novel connections between ranking and concepts from online stochastic optimization~\cite{agrawal2014fast} and model-predictive control~\cite{borrelli2017predictive}.
Furthermore, we evaluate the controllers on a number of synthetic and real-world datasets in Section~\ref{sec:experiments}, which provides practical guidance on when the use of each controller is most appropriate.

\subsection{Motivation 
\& Related Work} \label{sec:framework}
We argue that one of the key challenges in steering the long-term dynamics of AI platforms results from a mismatch in time scales. 
Algorithms on these platforms typically optimize metrics pertaining to individual requests or sessions, while the dynamics we aim to control play out over weeks or months of repeated interactions. 
Optimization on a per-request basis is ill-suited for even expressing long-term objectives, much less for steering their dynamics. 
Instead, we argue that we need a novel {\em macroscopic} view to enable strategic reasoning about the long-term dynamics of the platform in addition to the {\em microscopic} view that our methods currently focus on.

Existing work on incorporating long-term goals in search and recommendation has largely focused on fairness.
Early works defining fairness criteria posed them as constraints on impact or exposure to be fulfilled within a single ranking~\cite{singh2018fairness, zehlike2017fa,zehlike2017fa, celis2017ranking}.
Later work introduced a temporal perspective, including~\citet{celis2019controlling} who develop an online algorithm for recommending diverse viewpoints and~\citet{morik2020controlling,usunier2022fast} who present algorithms for satisfying fairness cumulatively over multiple rankings.
More recently, reinforcement learning algorithms have been applied to long-term fairness in order to handle endogenous dynamics---i.e., the impact of a ranking decision on future utilities and constraints~\citep{zhang2021recommendation, ge2021towards, yu2022policy}.
Beyond fairness, long-term exposure constraints have been motivated as an optimal strategy under the endogenous dynamics of content-provider viability~\cite{mladenov2020optimizing,zhan2021towards}.
Many of these settings fit into the framework that we propose.
However, unlike approaches which attempt to directly handle dynamics, we argue for elevating such strategic concerns to the definition of interventions. 

Partitioning into microscopic and macroscopic views has proven essential in the control of other complex systems. %
For example, macroeconomic metrics like gross domestic product and unemployment rate describe the state of our economy as a whole, and we use these metrics to reason about its long-term dynamics. 
Macro-level interventions are used to influence these metrics, like the interest rates set by the Federal Reserve. 
Similar examples are also widespread in engineered systems, where, for example, the macroscopic control intervention of a self-driving car (e.g., turn left 10 degrees) hides the microscopic execution of this command (e.g., voltages going to the steering motors) behind a control system.

\begin{figure}[t]
    \centering
    \begin{tikzpicture}%
    \node (start) [control] {Macro-Level Control of AI Platforms 
    
    \scriptsize \emph{Long-term sustainability of the platform.}\\
    Metrics: user satisfaction, supplier pool size, polarization, discrimination, ...\\
    Interventions: exposure allocation, diversification, novelty, external regulations, ...\\};
    \node (in1) [interface, below of=start, yshift=-0.5cm] {Macro/Micro Abstraction and Interface};
    \node (pro1) [optimization, below of=in1, yshift=-0.7cm] {Micro-Level Optimization of AI Platforms 
    
    \scriptsize \emph{Short-term utility maximization for participants.}\\
    Metrics: engagement through clicks, purchases, likes, streams, ...\\
    Interventions: ranking, artwork, push notifications, upsell, ...\\};
    
    \draw [arrow] (start) -- node[anchor=west] {\small Macro-level interventions} (in1);
    \draw [arrow] (in1) -- node[anchor=west] {\parbox{4cm}{\small Optimal \& consistent \\micro-level interventions}} (pro1);
    \end{tikzpicture}
    \caption{We propose to separate macro-level control used for steering the long-term dynamics of the platform from its micro-level engagement optimization. The interface layer provides an abstraction by optimally translating strategic macro-level interventions into a sequence of micro-level actions with minimal impact on short-term metrics.}
    \label{fig:micromacro}
\end{figure}
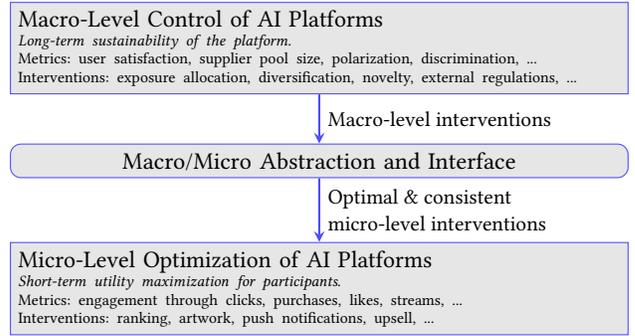

\fig{fig:micromacro} illustrates an analogous micro-macro view of AI platforms, where the macro-level metrics we aim to optimize are quantities like customer satisfaction, retention, polarization, or the size of the supplier pool. 
It is not hard to think of possible macro-level interventions either, like the rate with which the service is interrupting users with push notifications, how aggressively to prune clickbait, or how much exposure to give to smaller suppliers. 
Enabling reasoning at the level of macro-level metrics and interventions opens the door for future investigations of platform dynamics.
At the macro level, it will be far more tractable to understand how interventions affect the long-term metrics we aim to optimize for. 
For example, establishing a causal model of how exposure allocation to small suppliers (a scalar) relates to the size of the supplier pool (another scalar) is considerably less complex than estimating a causal link between millions of rankings and supplier-pool size.

In this paper, we focus on macro-level interventions that represent constraints on exposure or impact. 
Exposure can be quantified by models like the position-based model (PBM) \cite{craswell2008position} which assigns a score to each position in a ranking, representing the probability of being viewed by the user. 
The impact can be directly measured by clicks or purchases. 
Adding such long-term constraints provides a rich new language for guiding system behavior, as illustrated by the following examples that can be implemented in our framework:
    \begin{enumerate} \setlength\itemsep{-0.0mm}
        \item Give local artists at least $X$ percent of the overall exposure over the next month. (Item Group Exposure)
        \item Show new artist $A$ to at least $X$ users over the next week. (Single Item Exposure)
        \item Given a well-calibrated but imperfect spam filter, ensure that the expected exposure to spam across all users is less than $X$. (Item Group Exposure with uncertain Group Membership) 
        \item Do not send more than $X$ push messages on average per week to user $U$. (Single User Exposure)
         \item Show each housing-related ad to protected groups by demographic parity over the lifetime of the ad. (Single Item / User Group Exposure)
        \item Support goal of user $U$ to buy at least 30\% of products from local suppliers. (Single User Impact)
    \end{enumerate}
These examples show that macroscopic interventions can shape the aggregate experience of items over a given time span (first three), but also the aggregate experience of users (last three). 
The macroscopic interventions can provide constraints on the experience of a single user or item (2, 4, 6), on the collective experience of item or user groups (1, 3), or on the complex interaction between item and user groups (5). 
These interventions may be exact when we have precise knowledge of class membership, or they may be approximate and fulfilled only in expectation
(e.g., based on the probability of an article being spam). 
Finally, in some cases, the constraints directly act upon exposure (first five), while in example six the user asks the system to support a particular impact goal \cite{singh2018fairness} that includes the reactions of the users (e.g., clicks, purchases). Similar macro-level constraints are also relevant to other aspects of AI platforms like ads \cite{Yang/etal/19,Wu/etal/18a,Zhang/etal/22a,Zhang/etal/16} and ad-pacing \cite{Xu/etal/15,Agarwal/etal/14} in particular. 

\section{Macro/Micro Control Problem}
\label{sec:macro_micro_problem}

We now formalize the problem of translating a set of macro-level interventions into a sequence of micro-level actions.
We model this translation as a problem of optimal control which seeks to ensure that the micro-level actions achieve the desired macro-level interventions in aggregate. 
This is analogous to how control is used in mechanical systems to provide layers of abstraction with clear semantics. 
Controllers are used to keep a system in desirable states even under external perturbations and incomplete knowledge of the dynamics (e.g., keep a plane in level flight). 
Under bounds on worst-case conditions, controllers can be proven to be stable, safe, and performant \cite{Astroem/Murray/20,Sethi/Thompson/22}. 
Furthermore, the control perspective has already proven useful for other aspects of online systems \cite{Jambor/etal/12,Yang/etal/19,Xu/etal/15,Karlsson2013ApplicationsOF,Zhang/etal/16,control18jmlr}.

The macro/micro control problem takes the following form.
At each time step $t$ from $1$ to the final time step $\T$, a new context $\context_t$ arrives.
Each context $\context_t$ is drawn independently
from some unknown and possibly shifting sequence of distributions 
\begin{eqnarray*}
\context_t \sim P_t.
\end{eqnarray*}
At the micro-level, our goal is to derive a ranking policy that selects an action $\action_t$ (i.e., ranking) for each context that maximizes a micro-level ranking metric $\util(\action_t|\context_t)$  %
(e.g., Discounted Cumulative Gain (DCG) \cite{jarvelin2002cumulated}). 
Thus, over all contexts $\context_1, ..., \context_T$ our policy should choose actions $\action_1,...,\action_\T$ which achieve a large cumulative value:
\begin{eqnarray}
    \sum_{t=1}^T \util(\action_t|\context_t). \label{eq:microsum}
\end{eqnarray}
However, unlike conventional ranking policies, ours must also consider macro-level goals in addition to the micro-level utility. In particular, the policy needs to fulfill $\m$ constraints that range over all time steps from $1$ to $\T$:
\begin{eqnarray}
    \sum_{t=1}^{\T} \macrointer{1}(\action_t|\context_t) & \ge &  \macrotarget{1} \nonumber \\[-3mm]
    & \vdots & \\[-2mm]
    \sum_{t=1}^{\T} \macrointer{\m}(\action_t|\context_t) & \ge & \macrotarget{\m} \nonumber
\end{eqnarray}
Each of the $m$ constraints represents one macro-level intervention, where each $\macrointer{i}(\action_t|\context_t)$ represents a metric (e.g., exposure of an ad to demographic user groups, purchase of a local product), and each corresponding $\macrotarget{i}$ represents the cumulative goal over the time period from $1$ to $\T$. 
For a more compact presentation, we use the following vector notation to represent all $m$ constraints:
\begin{eqnarray}
    \sum_{t=1}^{\T} \macrointer{}(\action_t|\context_t) & \ge &  \macrotarget{} \:. \label{eq:macroconst} 
\end{eqnarray}
While maximizing Equation~\eqref{eq:microsum} subject to Equation~\eqref{eq:macroconst} may look like a straightforward optimization problem at first glance, it cannot be solved using standard optimizers. 
Crucially, contexts arrive sequentially, and the policy must immediately choose an action $\action_t$ at time step $t$  upon receiving $\context_t$. 
However, at time $t$, we only know the contexts $\context_1,...,\context_t$, but not the future contexts $\Context_t:=(\context_{t+1},...,\context_\T)$. This means that we cannot directly evaluate the constraints, and must pick the current action under partial information.

\begin{algorithm}[t]%
  \textbf{Input:} target $\macrotarget{}$, violation cost $\phi$, final time $\T$, controller $\policyclosed$ \\
  set initial state $\macrostate_0=0$ \\
  \ForEach{$t$ {\bf from} $1$ {\bf to} $\T$}{
     observe $\context_t \sim P_t$\\
     select action $\action_t = \policyclosed(\context_t,\macrostate_{t-1},t)$ \\
     update state $\macrostate_t = \macrostate_{t-1}+\macrointer{}(\action_{t}|\context_{t})$
  }
{\bf Compute ~objective:} $\sum_{t=1}^T \util(\action_t|\context_t) - \ViolCost^\top \hinge{\tau - \macrostate_\T}$
  \caption{Macro/Micro  Control Loop}
  \label{alg:closedloop}
\end{algorithm}

To address this problem, 
we first introduce a state $\macrostate_{t}$ that reflects the progress made towards fulfilling the constraints up to time $t$:
\begin{eqnarray}
    \macrostate_{t}=\sum_{t'=1}^{t} \macrointer{}(\action_{t'}|\context_{t'}).
\end{eqnarray}
Then, we consider ranking controllers of the form $\action_t=\policyclosed(\context_t,\macrostate_{t-1},t)$. 
The ranking $\action_t$ is chosen based on the context $\context_t$, the state $s_{t-1}$, and the time $t$.
The macro-level intervention is achieved when the constraints are fulfilled in the final time step $T$.
This is equivalent to ensuring that the terminal state $\macrostate_{\T} \ge \macrotarget{}$. 
The controller aims to reach this target state but must make decisions only on the basis of the current state and context without exact knowledge of the future contexts.

Fulfilling all constraints may not always be possible and, furthermore, it may not be desirable to arbitrarily sacrifice the micro-level objective from Equation~\eqref{eq:microsum}. 
We thus consider soft constraints and denote the \costterm~cost as
\begin{eqnarray}\label{eq:macro-violation-cost}
    \ViolCost^\top \hinge{\tau - \macrostate_\T},
\end{eqnarray}
where $\ViolCost \ge 0$ is an $m$ dimensional parameter vector that expresses how costly it is to violate each constraint. 
The ``hinge loss'' $\hinge{\cdot }$ sets all negative components of the input vector to zero, so any dimension of the terminal state $\macrostate_\T$ that is above its target $\tau$ contributes zero to the \costterm{} cost. 

The overall objective is the sum of the micro-level utilities minus the \costterm~cost at the final time step $\T$.
 For a given sequence of contexts, this objective is:
\begin{eqnarray}\label{eq:mainobjective}
    \sum_{t=1}^T \util(\action_t|\context_t) - \ViolCost^\top \hinge{\tau - \sum_{t=1}^T \macrointer{}(\action_t|\context_t)}\:.
\end{eqnarray}
Note that this final objective can only be computed after time step $\T$. Since the actions must be chosen sequentially,
this setting has the form of a closed-loop control problem, where the controller can react to the state $\macrostate_{t-1}$. This control loop is summarized in Algorithm~\ref{alg:closedloop}.

We conclude our general setup with a discussion of metrics which depend on \emph{modeled} vs. \emph{observed} feedback.
When metrics are defined by a model (e.g., the position-based model of exposure), the result of any arbitrary action $a$ can be anticipated once the context $x_t$ is observed.
In contrast, metrics defined by observed feedback (e.g., clicks or hover time) are known only for the chosen action $a_t$ and are observed only after the action is taken.
The control loop illustrated in Algorithm~\ref{alg:closedloop} is valid whether the metrics defining $u$ and $c$ are modeled or observed.
However, the action selection step defined by $\Pi$ benefits from the ability to anticipate the effect of arbitrary actions.
We, therefore, focus on modeled metrics in the controller development below---in other words, we take $u(\cdot|x_t)$ and $c(\cdot|x_t)$ to be known functions.
We call this the \emph{full information} setting, referring to the fact that the context provides sufficient information.
In practice, controllers may operate on the basis of imperfect models (e.g., using a learned relevance predictor instead of true relevances), even if the closed-loop logic proceeds according to observed feedback, and better models may be learned interactively by using this feedback. However, 
for clarity of exposition, we leave such a scenario to future work.

\subsection{Linear Utilities and Constraints}

We now define a specific class of models for describing the relationship between an arbitrary action $a$ and the macro- and micro-level objectives for a given context $x$.
For ease of exposition, we will refer to the micro-level objective $u(\action|\context)$ as ``utility'' and each macro-level $\macrointer{i}(\action|\context)$ as ``progress towards intervention $i$''.

\newcommand{\rk}{\mathrm{rank}}
The utility depends on both the relevance of the items and their ranked positions.
In the full information setting, the relevance of the items can be determined from context $x_t$.
Denote by $\mathbf r_{t,j}$ the relevance of item $j$ at time $t$.
Further define a position-dependent weight $\mathbf u_k$ for each position for $k\in[n]$ (e.g., Discounted Cumulative Gain \cite{jarvelin2002cumulated} is $\mathbf u_k=\log_2(k+1)^{-1}$).
Then an item $j$ ranked in position $k$ contributes $\mathbf r_{t,j} \mathbf u_k$ to the utility.
Denoting by $\rk(j| \action_t)$ the position of item $j$ under the ranking specified by $a_t$,
the utility has a linear form:
\begin{align*}
    \util(\action_t|\context_t) &= \sum_{j=1}^n \mathbf r_{t,j} \mathbf u_{ \rk(j | \action_t)}.
\end{align*}
Without loss of generality, we assume that $\mathbf u_k$ is non-increasing in position $k$, meaning that the higher an item is ranked, the more its relevance contributes to the utility. 

In the absence of macro-interventions, a utility-maximizing action sorts the items in order of their relevance scores: 
$a_t = \argsort \mathbf r_t$. However, our ranking controllers also consider progress towards the macro-level interventions of the following linear form:
\begin{align*}
    c_i(\action_t|\context_t) &= \sum_{j=1}^n    W_{t,ij}\mathbf e_{\rk(j\mid \action_t)} \quad i\in\{1,\dots,m\}\:. 
\end{align*}
Above,  $ W_{t,ij}$ determines the contribution of item $j$ to intervention $i$ (e.g., an indicator of group membership).
As with relevance, we assume the full information setting, so that the context $x_t$ contains enough information to determine this quantity.
Denoted by $\mathbf e_k$ is another position-dependent weight. 
It is not necessary to assume that it is equal to $\mathbf u_k$ or even that it is also non-increasing in $k$.

From here forward, we denote the parameters of the utility and progress functions by vectors and matrices: $\mathbf r_t\in\real^n$, $\mathbf u\in\real^n$, $ W_t\in\real^{m\times n}$, and $\mathbf e\in\real^n$.
We assume that position weights $\mathbf u$ and $\mathbf e$ are known\footnote{It is trivial to extend our controllers to the case that position-dependent weights vary with context, so long as the context provides full information on them.} and that the context provides full information about utility and interventions so that $x_t=(\mathbf r_t,  W_t)$.

For the ranking controllers that we develop, it is convenient to use permutation matrices for representing rankings.
A permutation matrix has exactly one entry equal to 1 in each row and column, and 0 elsewhere.
If $\Sigma\in\real^{n\times n}$ represents a ranking, then $\Sigma_{kj}=1$ means that item $j$ will be placed in position $k$.
Using this notation, the utility and exposure quantities can be written in a compact matrix-vector notation, identifying $a_t$ with the corresponding permutation matrix $\Sigma_t$:
\begin{align}
    \util(\action_t|\context_t) &=  \mathbf r_t^\top \Sigma_t \mathbf u,\quad 
    \macrointer{}(\action_t|\context_t) = W_t\Sigma_t \mathbf e.
\end{align}

Since searching in the discrete space of permutations can be computationally challenging, it is sometimes convenient to search over ranking distributions.
This corresponds to considering policies represented by doubly stochastic matrices rather than permutation matrices. The set of doubly stochastic matrices is defined as
\begin{align*}
    \Delta = \Big\{\Sigma \in \real_+^{n \times n}\mid  \sum_{k=1}^n\Sigma_{kj}=1~\text{and}~\sum_{j=1}^n\Sigma_{kj}=1~\forall~j,k\in\{1,\dots,n\}\Big\}.
\end{align*}
Given a doubly stochastic $\Sigma$, a ranking can be sampled via the Birkhoff-von Neumann decomposition \citep{birkhoff1946tres,singh2018fairness}.

\section{Controllers for Ranking}
\label{sec:controllers}
We now introduce three controllers to address the macro/micro control problem. 
We begin with a baseline \emph{myopic} controller that has a high reduction in micro-level engagement.
Next, we use the lens of online optimization to introduce a controller appropriate for \emph{stationary} context distributions and 
draw the connection to a previously proposed P-controller for ranking under fairness constraints \cite{morik2020controlling}.
Finally, we develop a more sophisticated \emph{predictive} controller 
that can anticipate and plan for non-stationarities in the context distribution.

\subsection{\basecontrollerLong~(\basecontrollerShort)}
Actions must be chosen at every time step $t$ without knowledge of future contexts.
As a result, the controller cannot exactly optimize \eqref{eq:mainobjective}.
A simple idea to address this issue is to define an intermediate objective at each time $t$:
\begin{eqnarray}\label{eq:myopicobj}
    \sum_{t'=1}^t \util(\action_{t'}|\context_{t'}) - \ViolCost^\top \hinge{ \frac{t}{T} \tau - \sum_{t'=1}^t\macrointer{}(\action_{t'}|\context_{t'})}\:.
\end{eqnarray}
This intermediate objective scales the target $\tau$ linearly by $\frac{t}{T}$
and removes the effect of future timesteps.
Note that this is not equivalent to the original objective
due to the nonlinearity of the hinge loss.
Effectively, this objective treats every timestep as if it were the final timestep (albeit scaling the target value).
Since there are no future timesteps to consider under this simplified objective, maximizing the objective for selecting the current action $\action_t$ is well-defined. This leads to the following controller, which we call the \basecontrollerLong{} (\basecontrollerShort{}):
\begin{align*}
   \policyclosed_\basecontrollerm(\context_t,\macrostate_{t-1},t) &= \argmax_{a} \util(\action|\context_t) - \ViolCost^\top  \hinge{\frac{t}{T}\tau - s_{t-1}-\macrointer{}(\action|\context_t)} \:.
\end{align*}
This expression contains only the terms from the objective~\eqref{eq:myopicobj} which affect the argmax. 
Notice that the maximizing action depends on the past actions and contexts only through the state $s_{t-1}$.
Algorithm~\ref{alg:basecontroller} presents the linear program (LP) implementation.

\begin{algorithm}[]%
  \textbf{Input:}  $x_t=(\mathbf r_t, W_t), s_{t-1}, t$\\
  \textbf{Parameters:}  $\tau\in\real^m$, $\ViolCost\in\real^m$\\
   $\Sigma_t = \argmax_{\Sigma \in \policyspace}  \mathbf r_t^\top \Sigma \mathbf u 
   - \ViolCost^\top  \hinge{\frac{t}{T}\tau - s_{t-1}-W_t \Sigma \mathbf e } $\\
    \textbf{Return:} $\action_t \sim \Sigma_t$\\
  \caption{\basecontrollerLong{} LP}\label{alg:basecontroller}
\end{algorithm}

While the intermediate objective at each time step is simple, it is overly strict. Specifically, it charges the full violation cost at the current time step if the controller is unable to reach the scaled target $\frac{t}{T}$. It thus ignores that the full violation cost is truly incurred only
in the final time step, and that the intermediate violations may cancel out before the final time step. We can thus expect this controller to perform very conservatively.

\subsection{Stationary Controller (\costreactiveShort)}

To address the inappropriate strictness of the Myopic Controller, we turn to ideas from online convex programming. Algorithms developed for this setting select optimization variables, in our case actions, at each time step based on streaming optimization parameters, in our case contexts~\cite{agrawal2014fast}.
As a first step, consider the following objective,
\begin{eqnarray}
    \min_{0\leq \lambda\leq \ViolCost} \frac{1}{T}\sum_{t=1}^T \util(\action_t|\context_t) - \lambda^\top \left(\frac{1}{T}\tau - \frac{1}{T}\sum_{t=1}^T \macrointer{}(\action_t|\context_t)\right),
\end{eqnarray}
where the inequality constraints on the Lagrange multiplier vector $\lambda\in\real^m$ are defined elementwise.
Besides rescaling by $\frac{1}{T}$, this is equal to the original objective~\eqref{eq:mainobjective}: minimizing over the multiplier $\lambda$ constrained to $[0,\phi]$ is an exact reformulation of the constraints implicit in the hinge loss appearing in the \costterm{} cost.

So far, the reformulation does not solve the problem of intermediate objectives, since minimizing the multiplier
$\lambda$ requires summing over the entire horizon.
However, if the value of $\lambda$ were fixed, 
then the objective would be separable over timesteps, and maximizing with respect to the action $a_t$ at time $t$ would no longer depend on the future. 
But we would have the same issues as the fixed violation cost in the \basecontrollerShort{}. 
How should a value of $\lambda$ be selected?
The key insight from the online optimization literature is to alternate between maximizing the objective while holding $\lambda$ fixed and updating $\lambda$ to iteratively minimize the objective.
Updating the multiplier in this way is like learning a dynamic violation cost.
Concretely, actions are chosen according to
\begin{align}
    a_t &= \argmax_a \util(\action|\context_t) - \lambda_{t-1}^\top \left(\frac{1}{T}\tau -  \macrointer{}(\action|\context_t)\right)\:.
\end{align}
This expression is simplified to contain only the terms  which affect the argmax.
It can be interpreted as approximating the average utility and macro-level progress over time as the utility or progress at each time step.
This is well motivated for i.i.d. contexts~\cite{agrawal2014fast}, and we therefore call this the \costreactiveLong{} (\costreactiveShort{}).

It remains to specify the multiplier updates.
In general, $\lambda_t$ is defined based on $\lambda_{t-1}$ and the gradient of the objective with respect to the multiplier: $\frac{1}{T}\tau - \macrointer{}(\action_t|\context_t)$.
In experiments, we use a variant of online gradient descent with adaptive step size.
For the sake of exposition, we derive a closed-form expression for the controller in the simpler case of gradient descent with fixed step size $\gamma>0$ and initialization $\lambda_0=0$. 
In this case,
\begin{align}
    \lambda_t &= \lambda_{t-1} + \gamma \left(\frac{1}{T}\tau - \macrointer{}(\action_t|\context_t) \right)
    = \gamma\left(\frac{t}{T}\tau - s_{t}\right).
\end{align}
The multiplier $\lambda_t$ is exactly the tracking error between the linearly scaled target and the current state, scaled by the step size $\gamma$.
Accounting for the bounds on $\lambda$, the closed-loop control law can be written as
\begin{align*}
	\policyclosed_\costreactivem(\context_t,\macrostate_{t-1},t) = \argmax_a  \util(\action|\context_t) +  \gamma\left(\tfrac{t-1}{T}\tau - s_{t-1}\right)_{[0,\ViolCost]}^\top  \macrointer{}(\action|\context_t).
\end{align*}
where $(z)_{[0,\ViolCost]}=\max\{0, \min\{\ViolCost, z\}\}$ is an elementwise clipping operation.
This closed-form expression illustrates that the \costreactiveLong{} adapts the weight on macro-level interventions depending on how much progress it has made towards the goal.
Algorithm~\ref{alg:reactcontroller} presents the linear program (LP) implementation.

\begin{algorithm}[]%
  \textbf{Input:}  $x_t=(\mathbf r_t, W_t), s_{t-1}, t$\\
  \textbf{Parameters:}  $\tau\in\real^m$, $\ViolCost\in\real^m$, $\gamma\in \real$,\\
   $\Sigma_t = \argmax_{\Sigma \in \policyspace}  \mathbf r_t^\top \Sigma \mathbf u - \left(\lambda_{t-1}^\top \right)_{[0, \ViolCost]}  W_t \Sigma \mathbf e  $\\
   update $\lambda_{t}$ from $\lambda_{t-1}$ with $\gamma$ and gradient $\frac{1}{T} \tau - W_t \Sigma_t \mathbf e$\\
    \textbf{Return:} $\action_t \sim \Sigma_t$\\
  \caption{\costreactiveLong{} LP}\label{alg:reactcontroller}
\end{algorithm}

\subsubsection{Proportional (P) control}

We briefly outline the connection between the \costreactiveLong{} and a (seemingly) 
heuristic method for boosting the position of certain items within a ranking.
Proportional (or simply ``P'') control is a general control technique which applies a correction proportional to the size of a tracking error.
In the context of ranking, P-control makes direct adjustments to relevance scores and was first proposed by~\citet{morik2020controlling} for achieving long-term fairness constraints.

To draw this connection, we write the control law $\policyclosed_\costreactivem(\context_t,\macrostate_{t-1},t)$ as a linear optimization problem and assume that $\mathbf u=\mathbf e$:
\begin{align}
\begin{split}
    \argmax_{\Sigma\in\Delta} \left(\mathbf r_t^\top + \gamma \left(\tfrac{t-1}{T}\tau - s_{t-1}\right)_{[0,\ViolCost]}^\top  W_t\right) \Sigma \mathbf u\\
    = \argsort \mathbf r_t + \gamma W_t^\top \left(\tfrac{t-1}{T}\tau - s_{t-1}\right)_{[0,\ViolCost]}
\end{split}
\end{align}
The equality holds because $\mathbf u$ is non-increasing.
Therefore, in this special case, the control law simply sorts items by adjusted relevance scores.
The adjustments are proportional to the tracking error,
where the transpose matrix $W^\top$
can be understood as translating from macro-level goals to individual items.
Therefore, items associated with lagging macro-level metrics will be boosted.
The step size parameter $\gamma$ can be interpreted as the ``gain'' of the P-controller, determining its sensitivity to tracking errors.

The P-controller is usually thought of as a heuristic, derived without reference to an overall objective.
Its score adjustment does not usually take into account the \costterm~cost parameter $\phi$
and it cannot account for possible differences between $\mathbf u$ and $\mathbf e$.
Despite this cost-obliviousness, the derivation above shows that P-control arises as a special case of our Stationary Controller.

\newcommand{\hatMacrointer}{\widehat{\Macrointer{}}}

\subsection{\costpredictiveLong{} (\costpredictiveShort)}

All controllers presented so far attempt to make progress towards macro-level goals at a constant rate over the horizon: at step $t$, the target is determined to be $\frac{t}{T}\tau$.
This does not account for non-stationarity in the context distribution.
For example, certain types of items may be relevant only on weekends or evenings.
Attempting to progress on macro-level goals at a linear rate fails to take into account variable underlying demand.

We therefore derive a predictive controller which accounts for the entire time horizon.
Denote future actions as
$\Action_{h}=(\action_{h+1}, \dots, \action_T)$ and similarly for contexts $\Context_{h}$.
The total progress can be written as
\begin{eqnarray}
    \sum_{t=1}^T
    \macrointer{}(\action_t|\context_t) =
    s_{h-1} + \macrointer{}(\action_h|\context_h) + \Macrointer{}(\Action_h|\Context_h) 
\end{eqnarray}
where $\Macrointer{}(\Action_h|\Context_h) $ is the ``progress-to-go'' at $h$ defined as the sum of $\macrointer{}(\action_t|\context_t)$ for $t$ from $h+1$ to $T$.
The first term in the above expression is the state: the accumulated progress so far.
The middle term is the contribution at time $h$, and the final term is the cumulative progress to come.
This expression explicitly separates the contribution of the past (known), present (current decision), and future (unknown).

The portion of the optimization objective~\eqref{eq:mainobjective} depending on the action $a_t$ at time $t$ can be written as
\begin{eqnarray}
    \util(\action_t|\context_t) - \ViolCost^\top \hinge{\tau - (s_{t-1} + \macrointer{}(\action_t|\context_t) +\Macrointer{}(\Action_t|\Context_t)  )}
\end{eqnarray}
Notice that because the utility is separable over time, the contributions of past and future utility do not impact the decision at time $t$.
Due to the hinge loss, the macro-level goal is not separable over time.
The progress-to-go $\Macrointer{}(\Action_t|\Context_t)$ depends on the future contexts, which are unknown, and future actions, which are hard to choose without knowing the contexts.
Instead, we propose using predicted values denoted by $\hatMacrointer_t$.
In Appendix \ref{app:forecast}, we present methods for forecasting the progress-to-go from historical data.

The following develops a multi-forecast predictive controller that can make use of such progress-to-go estimates.
Given a bootstrap sample of $B$ forecasts of the progress-to-go,
the predictive controller
selects an action which maximizes the average objective over these $B$ possible futures.
The multi-forecast objective at time $t$ is represented by the following optimization problem:
\begin{eqnarray}
    \max_a \frac{1}{B}\sum_{b=1}^B\util(\action|\context_t) - \ViolCost^\top \hinge{\tau - s_{t-1} - \macrointer{}(\action|\context_t) -\hatMacrointer_t^b}
\end{eqnarray}

\begin{algorithm}[b]
    \textbf{Input:}  $x_t=(\mathbf r_t, W_t), s_{t-1}, t$\\
  \textbf{Parameters:}  $\tau\in\real^m$, $\ViolCost\in\real^m$, $\gamma\in \real$, $\{(\hatMacrointer{}_t^b)_{t=1}^\top\}_{b=1}^B$\\
    $\Sigma_t =  \argmax_{\Sigma\in\Delta}  \mathbf r_t^\top \Sigma \mathbf u + \frac{1}{B}\sum_{b=1}^{B}  \left(\lambda_{t-1}^{b} \right)^\top_{[0, \ViolCost]}   W_t \Sigma \mathbf e$ \\
   update $\lambda_{t}^b$ from $\lambda_{t-1}^b$ with $\gamma$ and gradient $\macrotarget{} - \macrostate_{t-1} - W_t \Sigma_t \mathbf e - \hatMacrointer{}_t^b$ for $b=1,...,B$
   \\
 \textbf{Return:}~$\action_t \sim \Sigma_t$
\caption{\costpredictiveLong{} LP}\label{alg:predcontroller}
\end{algorithm}

Finally, we introduce multipliers for each forecast and use the same alternating online optimization approach as developed in the previous section for the \costreactiveLong{}.
Putting together all the pieces, the predictive controller 
$\policyclosed_\costpredictivem(\context_t,\macrostate_{t-1},t)$ selects actions according to:
\begin{align}
    a_t &= \argmax_a \util(\action|\context_t) + \frac{1}{B}\sum_{b=1}^B 
    \left(\lambda^b_{t-1}
    \right)_{[0,\ViolCost]}^\top \macrointer{}(\action|\context_t),
\end{align}
and updates each multiplier 
by defining $\lambda_t^b$ based on $\lambda_{t-1}^b$ and the gradient of the objective with respect to the multiplier: $\tau - s_{t-1} - \macrointer{}(\action_t|\context_t) - \hatMacrointer{}_t^b$ for $b\in[B]$.
For the simple case of online gradient descent, this update takes the form
\[    \lambda_t^b = \lambda_{t-1}^b + \gamma \left(\tau - s_{t-1} - \macrointer{}(\action_t|\context_t) - \hatMacrointer{}_t^b \right)\quad b=1,\dots,B.
\]
Similar to the \costreactiveLong{},  actions are chosen according to a weighted objective of utility and progress towards the macro-level targets.
However, while the \costreactiveLong{}\ updates the multiplier to target a linear rate of progress, 
the predictive controller updates the multiplier depending on potentially non-stationary forecasts of progress-to-go (e.g., expected higher demand on the weekend).
Algorithm~\ref{alg:predcontroller} presents the LP implementation.

\begin{figure*}[t!]
    \centering
    \includegraphics[width=.9\textwidth]{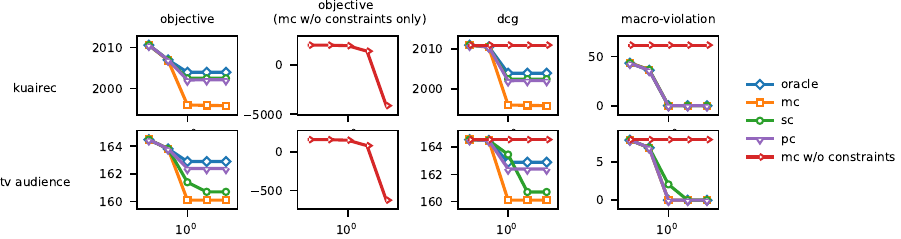}
    \vspace*{-1mm}
    \caption{Experiment results comparing all controllers across two datasets KuaiRec and Tv Audience. The x-axis is $\ViolCost$ on a log-scale in all plots. The first column is the final objective ~\eqref{eq:mainobjective} value, the middle column is the
    the utility metric (DCG), and the final column is the \costterm{}. The oracle has access to the test time contexts and directly optimizes the original objective~\eqref{eq:mainobjective}. The MC w/o constraints is an unconstrained utility maximizing controller.}
    \label{fig:main_figure}
\end{figure*}

\section{Experiments}
\label{sec:experiments}
While each controller comes with a strong conceptual and theoretical motivation, we now evaluate how far these arguments translate into improved empirical performance. 
In particular, we evaluate the controllers on real-world datasets to assess their differences on realistic data. Furthermore, we present experiments on synthetic data to explore in which situations \costpredictiveShort{} outperforms \costreactiveShort{}. 
Implementations of the controllers and code for reproducing the experiments are available 
at \url{https://github.com/xkianteb/ranking_constraints}.

\subsection{Experiment Setup}
In addition to the controllers discussed in Section~\ref{sec:controllers}, we include results for two additional controllers for comparison. As an (unachievable) skyline, we report the performance of an oracle controller that has access to the whole sequence of test time contexts and directly optimizes the overall objective. For further comparison, we also include the unconstrained controller (MC w/o constraints), which only optimizes utility without enforcing any macro-level interventions.

We conduct experiments on three datasets:
The first dataset is KuaiRec  \citep{gao2022kuairec}, which is a fully observed dataset collected from the recommendation logs of the video-sharing mobile app Kuaishou. 
The KuaiRec dataset consists of 1,411 users, 3,327 items, 4,676,570 interactions, and has a density of 99.6\%. 
We filter to include only items that every user has interacted with, which reduces the number of items from 3,327 to 2,062. 
We consider the task of ranking videos for sequentially arriving users.
Since the KuaiRec dataset does not provide relevance scores, we define the relevance score as half of the normalized watch ratio: $play\_duration / (2*video\_duration) $, capped at 1.
This is the recommended relevance signal provided by the dataset publishers \footnote{see the data description section on https://kuairec.com/}.
We define an exposure intervention on two arbitrarily chosen groups to evaluate the performance on multi-group constrained ranking. Each group contained two videos, one of which overlapped.
In particular, we set the exposure targets to be 1.1 times and 3 times the exposure of the unconstrained controller (MC w/o constraints).

The next dataset we consider is the linear television dataset Tv Audience \citep{turrin2014time}. 
This dataset contains temporal television watching behavior for 13k users: the watch duration consists of 217 channels over 19 weeks with an hourly time resolution.
For our experiment, we only use the first 12 weeks and ignored the remaining weeks, which results in a total of 288 timeslots. 
We consider the task of ranking channels over time.
The relevance score of a channel during a particular hour is defined as the number of viewers normalized by the channel's maximum viewers per hour over the past several weeks.
To evaluate the temporal prediction capabilities of controllers, we define an exposure intervention on a group of one arbitrarily selected late-night channel  that users mostly watch during the night or late evening hours.
The intervention is a 100\% exposure boost, which is equivalent to setting the exposure target to be twice that of the unconstrained controller (MC w/o constraints).

For our final dataset, we created a fully synthetic dataset to better understand the situations where \costpredictiveShort{} outperforms \costreactiveShort{}.
This dataset consists of temporal patterns of relevance scores for eight items over a horizon of 400 steps. 
The first four items are always relevant and have the highest relevance scores. 
The remaining four items are used to define exposure interventions of two disjoint groups. 
Two of the remaining items form one group and are most relevant during the first half of the time horizon, while the other two items form the second group and are most relevant during the second half of the time horizon.
Unlike the previous two datasets, all controllers are trained and evaluated on the same exact relevance scores. 
This means that we can assume accurate knowledge of the future context distribution when forecasting $\hatMacrointer_t$, which ensures that bad forecasts do not confound the evaluation of the controllers.

{\it Metrics. } For our experiments, we use discounted cumulative gain (DCG) \citep{jarvelin2002cumulated} as $\mathbf u$ for the utility metric and reciprocal rank (RR) 
$\mathbf{e}_k = 1/k$ as our exposure curve \cite{joachims2017unbiased}.

\begin{figure*}[t!]
    \centering
    \includegraphics[width=0.9\textwidth]{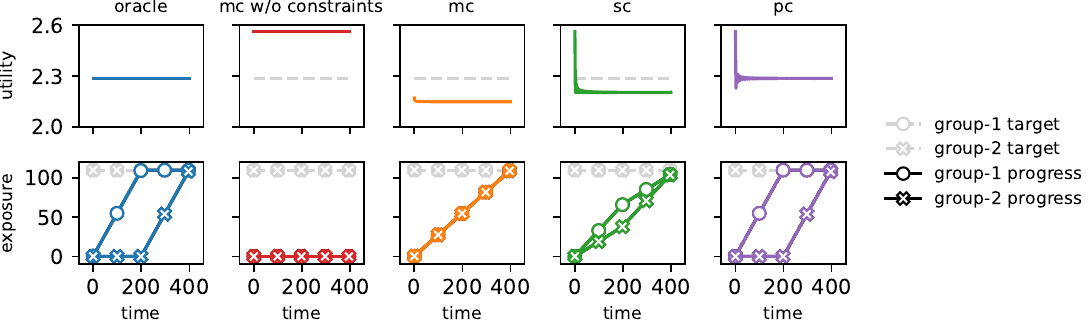}
    \vspace*{-1mm}
    \caption{Comparison of all controllers on a synthetic dataset to showcase when the \costpredictiveShort~ should be preferred. The o's and x's represent two groups of items. The top row is the average utility over time and the dashed grey line represents the highest achievable utility under the exposure constraint. The bottom row displays the exposure over time of both item groups. The grey x's and o's represent the target exposure for the groups.}
    \label{fig:counter_example}
\end{figure*}

{\it Hyperparameters. } 
\costreactiveShort{} and \costpredictiveShort{} have parameters that must be tuned and estimated to achieve good performance.
We tuned these parameters by dividing the data into three sets: train, development, and test.
For the \costpredictiveShort{}, we used the train set to estimate the forecast of the progress-to-go and the development set to simulate online contexts as described in Appendix~\ref{sec:hyperopt}.
We performed a grid search with the train and development sets to select the best forecast parameter based on the overall objective~\eqref{eq:mainobjective}. 
To update the multiplier $\lambda$, we utilized the Adam optimizer \citep{kingma2014adam} for both \costreactiveShort{} and \costpredictiveShort{}.
Additionally, we performed another grid search with the train and development datasets to pick the best Adam hyperparameters based on the overall objective~\eqref{eq:mainobjective}.
In our experiments, we investigated  the performance of each controller for different \costterm{} penalty values.
We selected the best-performing hyperparameters separately for each controller and penalty value,
and we repeated this process for all datasets. Detailed  hyperparameter ranges are given in Appendix \ref{app:expdetail}.

\subsection{Experiment Results}
In this section, we evaluate key properties of the controllers using the datasets 
discussed in the previous section. 
For all experiments, the x-axis represents the varying \costterm{} vector $\ViolCost$ (ranging from $1\times 10^{-2}$ to $1\times 10^{2}$), and the y-axis shows the performance of the controllers in terms of the utility~\eqref{eq:microsum}, violation cost~\eqref{eq:macro-violation-cost} or the overall objective~\eqref{eq:mainobjective}.

\subsubsection{Which controller achieves the highest overall objective?\\}

\cref{fig:main_figure} compares all controllers on both KuaiRec, a stationary dataset, and Tv Audience, a non-stationary dataset with temporal shifts in the context distribution. 
The performance of the oracle policy provides a skyline because it has (unrealistic) knowledge of all future contexts. 
We first note that for both datasets, all algorithms are largely equivalent when the \costterm{} cost factor $\ViolCost$ is small since a small cost factor implies little influence on the objective compared to short-term utility maximization.
As the importance of the long-term constraints increases with increasing \costterm{} cost factor $\ViolCost$, the \basecontrollerShort~ performs substantially worse than the other controllers, because its actions are chosen at every time step without consideration of how increasing violations in the current state interact with future contexts.
The right middle plots show, across both datasets, that under an increasing violation cost, all controllers eventually treat the macro-constraints as hard constraints and reduce their violation cost close to zero. However, both \costreactiveShort~ and \costpredictiveShort~ perform substantially better than \basecontrollerShort~ in terms of the overall objective. Comparing \costreactiveShort~ and \costpredictiveShort~ on the KuaiRec dataset, we see that both perform about the same. This is to be expected since there is no temporal pattern that \costpredictiveShort~ could learn to exploit with its progress-to-go estimates. Instead, it can only learn to predict the average exposure, which is precisely what the \costreactiveShort~ is optimizing. On the Tv Audience dataset, which has non-stationary temporal patterns, we see that the \costpredictiveShort~ can plan for the temporal pattern and perform better than all other controllers.

\subsubsection{How sensitive is \costpredictiveShort~ to the number $B$ of forecast samples?}
While the \costpredictiveShort~ performs well on temporal datasets and matches the performance of \costreactiveShort~ on non-temporal datasets, it has the number of forecast samples $B$ as an additional parameter that needs to be selected. \cref{fig:forecast} in the appendix shows the performance of \costpredictiveShort~ depending on the choice of $B$ for different cost vectors. Note that the results on Tv audience dataset are the median of 20 independent runs since the length of the test set is only 48 introducing a great amount of noise.  When the number of forecasts is greater than 20, then \costpredictiveShort~ performs well on both datasets, giving a reference point for selecting $B$. However, additional savings in computation time are possible for some datasets since we find that on the KuaiRec dataset, smaller values of $B$ can suffice to get good performance.

\subsubsection{ When is it advantageous to use a predictive controller?\\}
From the experiments in \cref{fig:main_figure}, we see that \costpredictiveShort~ performs better than other controllers on the Tv Audience dataset, which has a non-stationary temporal pattern. To further illustrate this point, \cref{fig:counter_example} is a salient example that explores when the \costpredictiveShort{} should be preferred over other controllers that are oblivious to any temporal patterns. 
In the figure, each column represents the controller utility at the top and exposure at the bottom. There are eight items in total and one item in each group. We compute nDCG@4 and RR@4; which means only the top-4 items receive utility and exposure. When no macro-level exposure goals are enforced, as seen in the MC w/o constraints, the exposure for both groups is zero because they are not among the top-4 ranked items. 
As illustrated by the oracle controller, the first group is more relevant during the first half of the time horizon, while the second group is more relevant during the second half.
The \costpredictiveShort{} performs similarly to the oracle controller because it can leverage this temporal pattern. 
The \costreactiveShort{} is better compared to \basecontrollerShort{}
due to tuning the gain parameters which allows for additional flexibility. 
However, both \basecontrollerShort{} and \costreactiveShort{} follow a linear exposure target and are thus unable to boost the two groups separately.

\section{Conclusions}
We formalize and address the problem of how to design algorithms that convert macro-level goals into a sequence of individual rankings that have the least impact on micro-level metrics. The algorithms we introduce are analogous to how control is used in mechanical systems to provide layers of abstraction with clear semantics. By introducing three new controllers, we cover a range of application scenarios. Furthermore, we provide rigorous justification for proportional controllers for ranking. Of the three controllers we introduce, we find that controllers based on online optimization  (i.e., \costreactiveShort~ and \costpredictiveShort) outperform the more naive \basecontrollerShort~ controller. Furthermore, we find that the predictive controller (\costpredictiveShort) performs better than the stationary controller (\costreactiveShort) in non-stationary settings. 

This paper opens up a wide range of future work. By making new technical connections between ranking and control theory, it opens up a new set of tools for designing adaptive ranking policies. Furthermore, we anticipate that the macroscopic view of ranking platforms we introduce will provide a conceptual framework for making these platforms more steerable.

\section{Acknowledgments}

This research was supported in part by NSF Awards IIS-2312865, IIS-2008139, and CCF 2312774; a CI Fellowship; and a gift from Wayfair. We thank Ali Vanderveld and Benjamin Schroeder for their valuable input. All content represents the opinion of the authors, which is not necessarily shared or endorsed by their respective employers and/or sponsors.

\newpage

\section{Ethical Considerations}
Understanding that search and recommendation AI platforms are socio-technical systems means that platform designers must consider the systems' technical and social aspects. In particular, when recommendation systems are being optimized for short-term engagement, it is potentially at the cost of the long-term sustainability of the platform. Not considering long-term sustainability could have social implications like amplifying misinformation or providing disparate utility to different groups. This could affect users' long-term satisfaction with a platform. Our work focuses on developing algorithms that incorporate a platform designer's long-term goals while maintaining a platform's short-term goals. Our intent is that this will enable platform designers to better incorporate all of the socio-technical aspects of a system into the algorithmic decision-making of a search and recommendation AI platform. However, the increased ability to control long-term platform behavior does not automatically lead to better behavior, and responsible governance in setting long-term goals is of crucial importance.

\bibliographystyle{ACM-Reference-Format}
\balance
\bibliography{acmart,ref}

\clearpage
\newpage

\appendix
\section{Notation Summary}

\begin{align*}
  &t && \text{time}\\
  &T && \text{max time}\\
  &n && \text{number of items}\\
  & \mathbb{P}=(P_1, P_2, \dots, P_T) &&  \text{nonstationary context distribution}\\
  &\context_t\sim P_t && \text{context}\\
  &\action_t && \text{ranking at time}~t\\
  &\Context_t \coloneqq (\context_{t
  +1}, \dots, \context_{T}) && \text{future contexts at time}~t\\
  &\Action_t \coloneqq (\action_{t+1}, \dots, \action_{T}) && \text{future rankings at time}~t\\
& \cline{1-2} \\
  &\util(\action|\context) \in\real && \text{micro-metric}\\
  & \m && \text{number of interventions}\\
  &\macrotarget{} \in \real^{m} && \text{macro-target}\\
  &\macrointer{}(\action|\context)\in\real^\m && \text{macro-metric}\\
  &\macrostate_t = \sum_{t'=1}^{t} \macrointer{}(\action_{t'}|\context_{t'})  && \text{progress state}\\
  & \ViolCost \in \real^{m} && \text{\costterm{} cost vector}\\
  &\macroInter(\Action_t|\Context_t) = \sum_{t'=t+1}^{T} \macrointer{}(\action_{t'}|\context_{t'})  && \text{progress-to-go at}~t\\
  & \cline{1-2} \\
  & \mathbf r_t\in\real^n && \text{relevance scores}\\
  & W_t \in\real^{m\times n} && \text{intervention-item association}\\
  & \textbf u, \textbf e\in\real^n && \text{micro and macro position weights}\\
  & \Sigma\in\Delta && \text{doubly stochastic ranking matrix} \\
\end{align*}

\section{Controller Implementation}
The controllers developed in Section~\ref{sec:experiments} depend on various parameters. All depend on the macro-level intervention targets $\tau$ and violation cost vector $\ViolCost$, which we assume are specified by designers. 
The \costreactiveLong{} and \costpredictiveLong{} additionally depend on an optimization parameter $\gamma$ which determines the multiplier updates.
Due to the connection with P-control discussed above, we refer to this parameter as the \emph{gain}.
Additionally, the \costpredictiveLong{} depends on forecasts of the progress-to-go. 
We now discuss how to use offline data to determine these quantities.

\subsection{Estimating the Progress-to-go} \label{app:forecast}
The \costpredictiveLong{} requires several forecasts of the progress-to-go. 
We estimate these forecasts using offline data.
The offline data defines an empirical distribution of contexts over time.
Suppose that the offline data contains $N$ contexts which we will index by $j$.
Then for offline bootstrap sample $b$ and time step $t$ we sample the context by its index $j_{bt}$.
We do so in a stratified manner to preserve relevant temporal relationships in the data.
Depending on the setting, this sampling procedure may treat hour of the day, day of the week, etc, equivalently.
By sampling contexts by their indices for all $t$ and $b$, we construct several sampled sequences of contexts from this time-dependent distribution: $\{(x_{j_{bt}})_{t=1}^T\}_{b=1}^{B_\mathrm{off}}$.

However, the progress-to-go is not determined only by contexts---it also depends on the sequence of actions.
We construct actions using an offline optimization over the entire horizon, using the exact objective~\eqref{eq:mainobjective} and the sampled sequences of contexts.
Algorithm~\ref{alg:forecast} presents the resulting LP.
Notice that the ranking actions are not independent at each time step or in each bootstrap sample.
Rather, the optimization problem finds the best contextual policy that is stationary over time.
This formulation has two advantages.
First, it prevents the optimization problem from growing with the horizon or with the number of bootstrap samples.
Second, by constraining the actions to come from this reduced policy class, it prevents the offline optimization problem from over-exploiting its ability to see all contexts, in contrast to the partial information faced by the online controller in practice.
This helps to ensure that the forecasted progress-to-go variables are not too ambitious.

Finally, the resulting sequence of (approximately) optimal actions 
along with the sampled context sequence define the progress-to-go at each time step.
A total of $B_\mathrm{on}\leq B_\mathrm{off}$ forecasts are created through this process.

\begin{algorithm}
   \textbf{Input:} Dataset defining an empirical distribution $\widehat{\probSet}$\\
    
    \ForEach{$t=1,\dots,T$,~~$b=1,\dots,B_\mathrm{off}$}{
    Sample context $x=(r,W)\sim \widehat{\probSet}$ at index $j_{bt}$.
 }

    \begin{align*}\widehat{\boldsymbol{\Sigma}} = \argmax_{(\Sigma_1, \dots, \Sigma_N)\in\Delta}  \frac{1}{B_\mathrm{off}}\sum_{b=1}^{B_\mathrm{off}} \Bigg[ &\sum_{t=1}^{T} r_{j_{bt}}^\top \Sigma_{j_{bt}}\mathbf u \\
    &- \phi^\top  \Big(\tau - \sum_{t=1}^{T} W_{j_{bt}}\Sigma_{j_{bt}}\mathbf e \Big)_+ \Bigg]
    \end{align*}\\

 \ForEach{$t=1,...,T$ and $b=1,...,B_\mathrm{on}$}{
 $\hatMacrointer_t^b =   \sum_{t'=t+1}^{\T}   W_{j_{bt'}} {\widehat{\Sigma}_{j_{bt'}}}e $ \\
 }
\caption{Forecasting Progress-to-go}\label{alg:forecast}
\end{algorithm}

\subsection{Tuning the Gain} \label{sec:hyperopt}
Both the \costreactiveLong{} and \costpredictiveLong{} depend on the gain parameter $\gamma$.
To tune its value, we again use offline data to sample sequences of contexts.
We sample contexts in a time-sensitive manner as described in the previous subsection.
These samples are used to simulate closed-loop control and the resulting performance approximates the performance of a controller with the given parameter $\gamma$. 
We then do a simple grid search on the parameter $\gamma$.
Algorithm~\ref{alg:tuning} describes this procedure.

\begin{algorithm}%
    \textbf{Input:} \\
    \ForEach{$t=1,\dots,T$}{
    Sample context $x_t=(r_t,W_t)\sim \widehat{\probSet}$.
    }
    \textbf{Initialize:} $\gamma_{best}$,~ $\Loss_{best}=-\infty$,~$s_0=0$\\
    \For{$\gamma \in \{ 10, 1, .1, .01, etc. \}$}
    {
     \ForEach{$t=1,\dots,T$}{
        play~$a_t =\policyclosed_\gamma(\context_t,\macrostate_{t-1},t)$ \\
        $s_{t} = s_{t-1} + \macrointer{}(\action_t|\context_t)$\\
        }
        $L = \sum_{t=1}^\top \util(\action_t|\context_t) - \ViolCost^\top \hinge{\tau - s_T}$\\
        \If{$\Loss \geq \Loss_{best}$}{
            $\Loss_{best} = \Loss$,~$\gamma_{best} = \gamma$\\
        }
    }
    \textbf{Return:} $\gamma_{best}$
    \caption{Tuning Loop}\label{alg:tuning}
\end{algorithm}

\subsection{Multiplier Update Algorithms}

There are many possible ways to update the multiplier variables in implementing \costreactiveShort{} and \costpredictiveShort{}.
Perhaps the simplest is Online Gradient Descent (Algorithm~\ref{alg:ogd}).
We initially experimented with Gradient Descent with Momentum, but
ultimately we found the Adam optimizer \citep{kingma2014adam} (Algorithm~\ref{alg:adam}) to be most successful.

\begin{algorithm}[]%
  \textbf{Input:}  $\lambda_{t-1}$, gradient $g_t$\\
  \textbf{Parameters:}   $\gamma\in \real$\\
    \textbf{Return:} $\lambda_t = \lambda_{t-1} - \gamma g_t$\\
  \caption{Online Gradient Descent Update}\label{alg:ogd}
\end{algorithm}

\begin{algorithm}[]%
  \textbf{Input:}  $\lambda_{t-1}$, gradient $g_t$\\
  \textbf{Parameters:}   $\gamma\in \real$, $\beta\in \real$, $\epsilon\in \real$,  $m_0=0$, $v_0=0$\\
  \textbf{Update:} \begin{eqnarray*}
    &m_t=\beta m_{t-1} + (1-\beta) g_t 
    &v_t=\beta v_{t-1} - (1-\beta) g_t^2\\ 
    &\hat m_t=m_t/(1-\beta^t)\quad  &\hat v_t=v_t/(1-\beta^t) \end{eqnarray*}
    \textbf{Return:} $\lambda_t = \lambda_{t-1} - \gamma {\hat m_t}/{\sqrt{\hat v_t+\epsilon}}$\\
  \caption{Adam Update}\label{alg:adam}
\end{algorithm}

\begin{table}
\centering
   \begin{tabular}{l|l|l}
       \toprule
       Controller & Hyperameter & Values \\
       \midrule
       \costreactiveShort & Adam-$\beta$ & 0.5, 0.9, 0.98 \\
                          & Adam-$\epsilon$ & 1e-05, 1e-08 \\
                          & Optimizer & Adam \\
                          & gain $\gamma$ (learning rate) &  1e-3, 1e-2, 1e-1, 1e0, 1e1,\\
                          &  & 1e2, 1e3  \\
                                 \midrule
       \costpredictiveShort & Adam-$\beta$ & 0.5, 0.9, 0.98 \\
                          & Adam-$\epsilon$ & 1e-05, 1e-08 \\
                          & Optimizer & Adam \\
                          & gain $\gamma$ (learning rate) &  1e-3, 1e-2, 1e-1, 1e0, 1e1, \\
                          & & 1e2, 1e3  \\
                          & \# offline-forecast & 20, 50 \\
                          & \# online-forecast & 20, 50 \\
       \bottomrule
   \end{tabular}
   \vspace{0.5mm}
   \caption{Hyperparameters used for experiments}
   \label{tbl:imdb_hparams}
\end{table}

\newpage

\section{Additional Experiment Details} \label{app:expdetail}

\subsection{Hyperparameters}

We provide a table the hyperparameters that we used for tuning each of the controllers in our experiments.

\subsection{Additional Experiments}
We conduct additional ablation experiments using the Last.fm dataset \citep{celma2010lastfm}. This dataset contains tuples of users, artists, and the amount of time that a user listened to a particular artist. 
We consider the task of ranking artists for sequentially arriving users.
We define the relevance score for a user and artist to be the play time and consider a subset of artists. 
Of the total 292,385 artists and 358,868 users, we perform experiments using a subset of 1,373 users and 50 artists. 
We define an exposure intervention on a group containing two artists.
The two artists are selected based on the listening behavior of two disjoint sets of users.
Within each set of users, the top-15 artists are similar, but between the two sets, they are non-overlapping.
The two artists are chosen to be the 15th most popular artist within each of the two user sets
and the exposure target is set to ten times the original unconstrained maximizing ranker exposure.
We use this structure to create a temporal pattern in the data: for the first half of the time steps, contexts are defined by users sampled from the first set of users. 
In the second half, they are sampled from the second set.
Furthermore, for this dataset, we only consider DCG@15 instead of DCG across
the entire set of items to be ranked.

\begin{figure}[h!]
    \centering
    \includegraphics[width=.9\columnwidth]{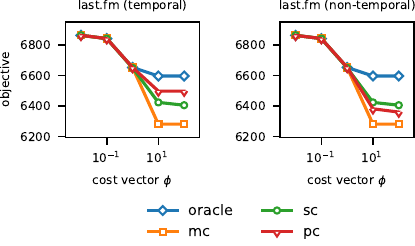}
    \caption{Comparison of two different versions of the last.fm dataset. The left plot enforces a temporal pattern during training, and the right plot shuffles the dataset and breaks the temporal pattern. Furthermore, the test time contexts have a temporal pattern, and the target exposure is kept the same across both plots.}
    \label{fig:lastfm}
\end{figure}

\begin{figure}[h!]
    \centering
    \includegraphics[width=.8\columnwidth]{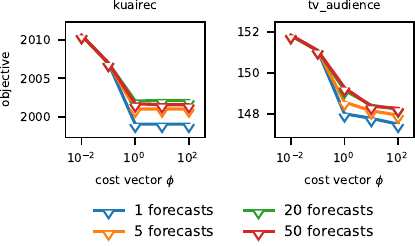}
     \caption{Comparison of different forecast samples used for computing the progress-to-go. The KuaiRec dataset on the left is non-temporal and the Tv Audience dataset on the right is temporal.}
    \label{fig:forecast}
\end{figure}

\subsubsection{Does removing the temporal pattern affect performance?}
\cref{fig:lastfm} shows performance on the two versions of the Last.fm setting. For a fair comparison we keep the dataset and splits the same and only vary shuffling order. The non-temporal version of last.fm shuffles the contexts, which breaks the temporal pattern. We see that similar to the $\cref{fig:main_figure}$, when the cost is small, all controllers trade of small violation to obtain a larger objective value. We see that \costpredictiveShort{} performs worse when using the non-temporal dataset than the temporal dataset.

\end{document}

%% file: macros.tex
\usepackage{xcolor,colortbl}
\usepackage{hyperref}
\usepackage[ruled,vlined]{algorithm2e}
\usepackage{bm}

\newcommand{\basecontrollerShort}{MC}
\newcommand{\basecontrollerLong}{Myopic Controller}
\newcommand{\basecontrollerm}{\mathrm{MC}}

\newcommand{\costreactiveShort}{SC}
\newcommand{\costreactiveLong}{Stationary Controller}
\newcommand{\costreactivem}{\mathrm{SC}}

\newcommand{\costpredictiveShort}{PC}
\newcommand{\costpredictiveLong}{Predictive Controller}
\newcommand{\costpredictivem}{\mathrm{PC}}

\newcommand{\costterm}{macro-violation}

\DeclareMathOperator*{\argmax}{argmax}

\newcommand{\Loss}{L}               %
\newcommand{\util}{u}                    %

\newcommand{\m}{{m}}

\newcommand{\argsort}{\operatornamewithlimits{argsort}}

\newcommand{\real}{\mathbb{R}}

\newcommand{\hinge}[1]{{\left({#1}\right)_+}}  %

\newcommand{\fig}[1]{Figure~\protect\ref{#1}}

\newcommand{\context}{x}             %
\newcommand{\Context}{X}             %
\newcommand{\action}{a}              %
\newcommand{\Action}{A}              %

\newcommand{\T}{T}                   %

\newcommand{\ViolCost}{\phi}            %

\newcommand{\Macrointer}[1]{C_{#1}} %
\newcommand{\macrointer}[1]{c_{#1}} %
\newcommand{\macroInter}{C}
\newcommand{\macrotarget}[1]{\tau_{#1}} %
\newcommand{\macrostate}{s}

\newcommand{\policyspace}{{\Delta}}         %

\newcommand{\policyclosed}{\Pi}           %

\newcommand{\probSet}{\mathbb{P}}

\newtheorem{User Model}{User Model}

\usepackage{tikz}
\usetikzlibrary{shapes.geometric, arrows}

\tikzstyle{control} = [rectangle, 
text width=8cm, 
draw=blue!70, fill=gray!20]

\tikzstyle{interface} = [rectangle, rounded corners, 
text width=8cm, 
text centered, 
draw=blue!70, fill=gray!20]

\tikzstyle{optimization} = [rectangle, 
text width=8cm, 
draw=blue!70, fill=gray!20]

\tikzstyle{arrow} = [thick,->,>=stealth,draw=blue!70]